# Interacting many-body systems in quantum wells: Evidence for exciton-trion-electron correlations

M. T. Portella-Oberli, V. Ciulin*, J. H. Berney, and B. Deveaud
*Institut de Photonique et d'Electronique Quantiques - Ecole Polytechnique Fédérale de Lausanne (EPFL) – CH1015 Lausanne, Switzerland*

M. Kutrowski , and T. Wojtowicz
*Institute of Physics, Polish Academy of Science, aleja Lotnikow 32/46, 02-668, Warsaw, Poland*

We report on the nonlinear optical dynamical properties of excitonic complexes in CdTe modulation-doped quantum wells, due to many-body interactions among excitons, trions and electrons. These were studied by time and spectrally resolved pump-probe experiments. The results reveal that the nonlinearities induced by trions differ from those induced by excitons, and in addition they are mutually correlated. We propose that the main source of these subtle differences comes from the Pauli exclusion-principle through phase-space filling and short-range fermion exchange.

PACS: 71.35.Pq, 78.47.+p, 42.65.-k, 78.67.De

Many-body interactions are the main source of nonlinear optical properties in condensed matter physics, and more particularly in semiconductors. In undoped semiconductors, Coulomb interaction gives rise to the electron-hole bound states (excitons) that play a crucial role in determining the optical properties near the band edge. Exciton resonances show a number of interesting features, both in the linear and in the nonlinear regimes making their study a very active field of research over the last 20 years [1]. The nonlinear dynamical properties of excitons are a very direct probe of the many-body interactions occurring in quantum wells, which might eventually find some useful applications in optoelectronic. The optical spectrum of moderately doped quantum wells also features a trion (charged exciton) resonance, which is situated only few meV below the exciton line [2, 3]. The way many-body interactions are modified by the presence of an additional electron gas is clearly a topic of major interest for applications such as transport of light by a charged exciton [4] and for the quantum-information science [5]. Current proposals for a solid-state implementation of an all-optical spin-based quantum computer are relying on Coulomb interaction between electrons and excited excitons either via electron exchange due to electron-exciton interaction [6] or via electron-charged excitonic states (trions) [7]. Modulation-doped quantum well provides a model system in which electrons, excitons and charged excitons cohabit in the same well. The many-body interactions among electrons excitons and trions may be probed through the non-linear behaviour of trion and exciton optical resonances.

In this paper, we report for the first time, the experimental results on the dynamical nonlinear optical properties of trions and excitons in modulation-doped quantum wells. We evidence correlated behavior of excitons and trions under excitation which manifests itself by crossed trion-exciton effects. We observe a wealth of phenomena encompassing bleaching, crossed bleaching, induced-absorption and energy shifts of the resonances. Significant differences are found between the nonlinear optical effects induced by an exciton and a trion population. The main source of these distinct differences is proposed to come from the Pauli exclusion-principle, which is at the origin of phase-space filling and short-range fermion exchanges.

We use here a high quality sample that was already fully characterized in previous studies [8-11]. The sample is a one-side modulation-doped $CdTe/Cd_{0.73}Mg_{0.27}Te$ heterostructure containing one single quantum well of 8 nm. Five monolayers of iodine were incorporated into the barrier, separated from the quantum well by a spacer of 10 nm [8, 11]. The presence of iodine donors results in an electron concentration in the well of about $4\times10^{10}$ $cm^{-2}$. In Fig. 1, we plot the measured and calculated linear reflectivity spectra of the sample. The latter was calculated using transfer matrix model [12] and we have obtained by this method a spectrum very similar to the one we measure. The shape of the reflectivity spectrum is due to the sample structure. In our sample the measured and calculated reflectivity spectra have absorption like line shape. This is because the quantum well is situated at a distance close to $\lambda/4$ from the surface (the cap layer thickness is 50 nm). The reflectivity spectrum would be dispersive for cap layers with other thickness. Our sample has been grown with precisely defined parameters to obtain the desired linear reflectivity spectrum that allows probing absorption features from the reflectivity spectrum. This reflectivity spectrum (Fig. 1) evidences a spectral line situated about 3 meV below the neutral exciton. This line is assigned to negatively charged excitons, trions [3, 8, 10, 11]. The exciton and the trion resonances were unambiguously identified as they follow the adequate selection rules under magnetic field [3, 11] and a transfer of the line intensities from exciton to trion is observed with increasing electron concentration [10, 11]. Therefore, the exciton and trion resonances are very well defined to be probed in pump and probe experiments. All the experiments described in the present paper were also performed on a less doped part of the sample (electron gas concentration of about $2\times10^{10}$ $cm^{-2}$) and the results obtained show a similar behaviour.

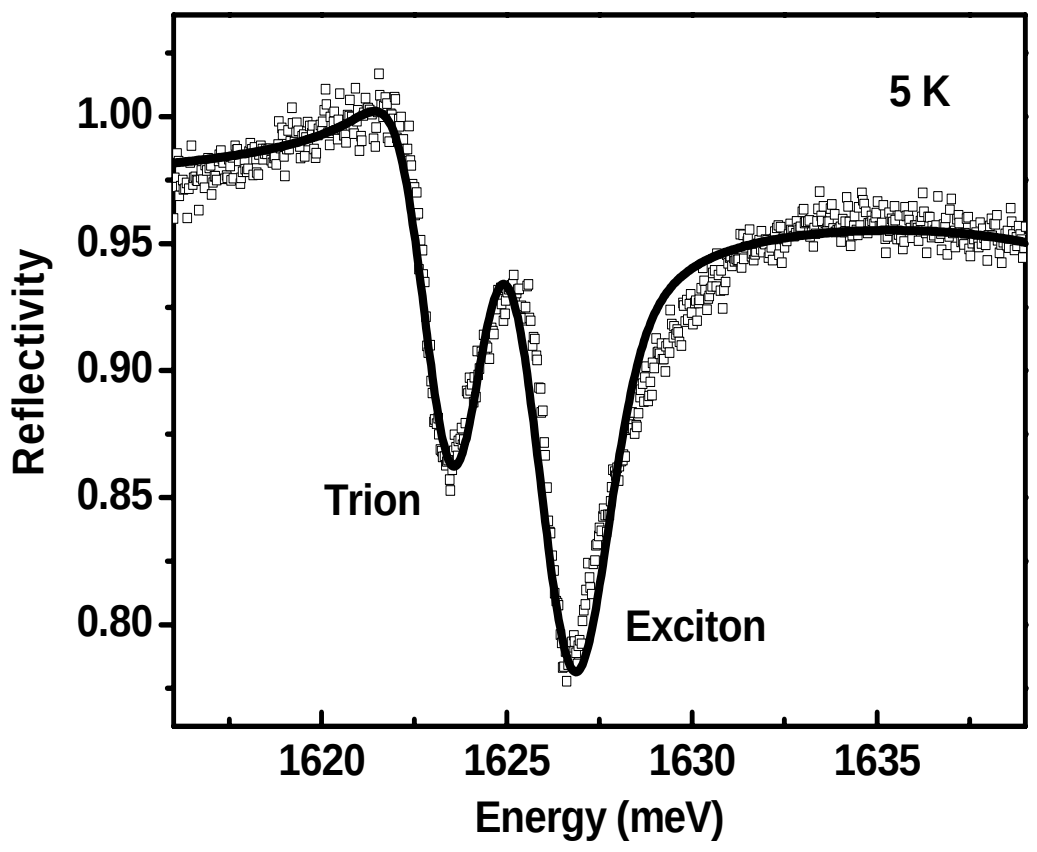

**FIG. 1:** Measured (square symbol) and calculated (solid line) reflectivity spectra of the sample at 5 K.

Due to the similarity between reflectivity and absorption in our sample, the absorption changes can be directly quantified on the reflectivity spectrum. This is confirmed by a simple calculation, using again the transfer matrix formalism, showing that a variation of the trion and exciton oscillator strength ($\Delta f_T/f_T$ and $\Delta f_x/f_x$, respectively) gives rise linearly to a variation in the differential reflectivity ($\Delta R_T/R_T$ and $\Delta R_x/R_x$, respectively). For 1 % of $\Delta f_T/f_T$ and $\Delta f_x/f_x$ has been quantified precisely 1.2 % and 1.6 % of $\Delta R_T/R_T$ and $\Delta R_x/R_x$, respectively. Therefore, to attain suitable information about nonlinear effects in the excitonic complexe resonances, we can use the differential reflectivity spectrum obtained through picosecond-pump and femtosecond-probe experiments. For these measurements, the laser beam from a titanium-sapphire laser is split into two parts: one passes through a pulse shaper to generate a narrow tunable 1.3 picosecond pulse and the other one, the broad 100 femtosecond pulse, about 8 times less intense, is sent to a delay line, both beams are focused onto the sample. The two laser beams were cross-linearly polarized. The wavelength of the narrow pump pulse is varied systematically to excite specific energies across the exciton and trion resonances, and the excitation density is varied typically from $10^9$ cm$^{-2}$ up to $10^{10}$ cm$^{-2}$. The modifications in the optical spectrum are probed by the broad probe pulse dispersed by a 1 m monochromator and detected by a cooled ccd. As the changes induced by the pump to the linear spectra are weak and very wavelength dependent, we have measured the differential reflectivity spectrum $\Delta R = (R-R_0)/R_0$ for each delay time between the pump and probe pulses, $R_0$ and R being the unexcited and excited reflectivity spectra, respectively.

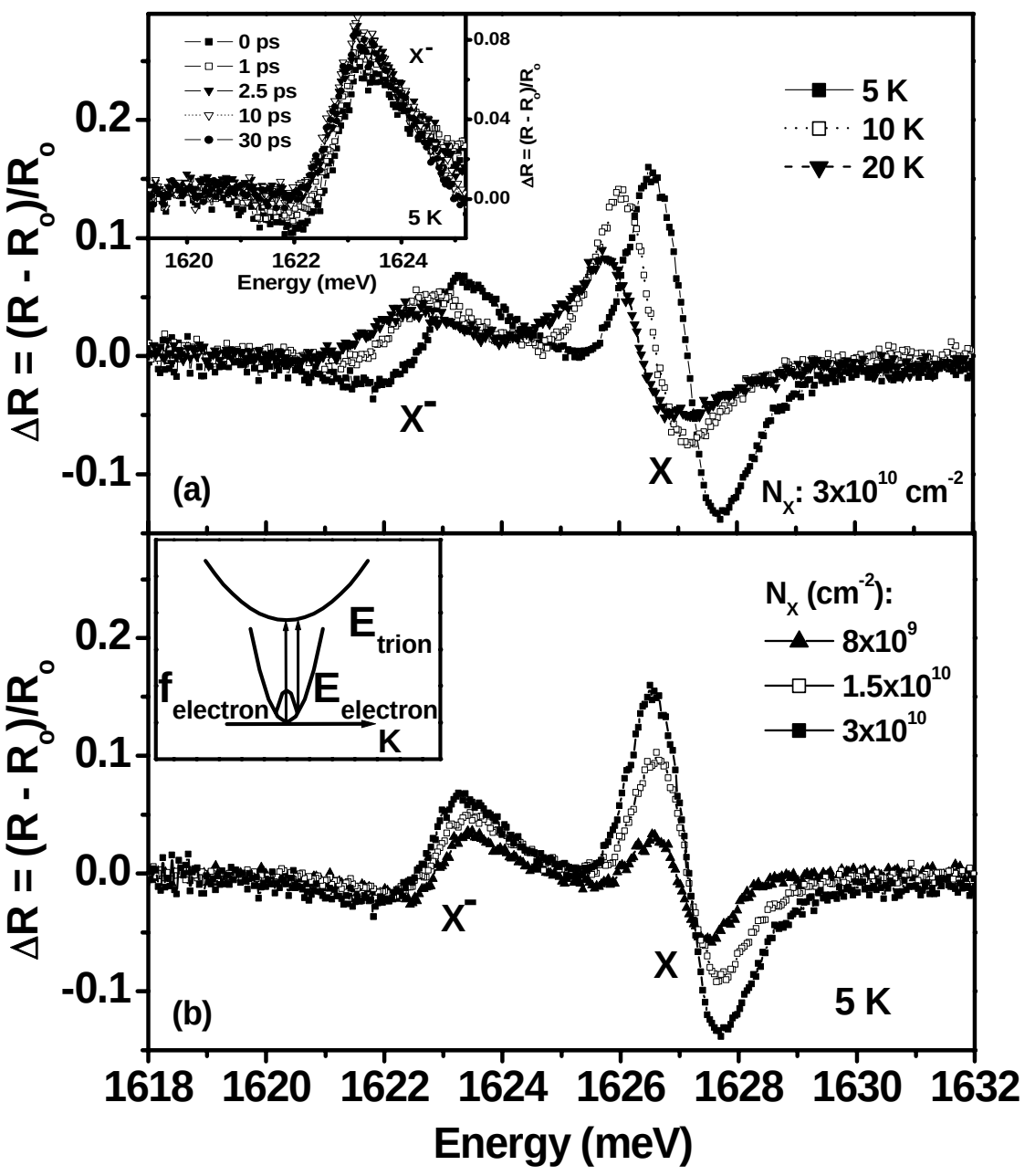

**FIG. 2:** Differential reflectivity spectra obtained by pumping at the exciton resonance, at zero delay time between pump and probe pulses: **(a)** at temperatures of 5 K and excitation energies $E_P$ = 1626.65 meV, at 10 K and $E_P$ = 1626.4 meV and at 20 K $E_P$ = 1626.0 meV and , **(b)** at 5 K, $E_P$ = 1626.65 meV and exciton densities of about: 8x$10^9$ cm$^{-2}$, 1.5x$10^{10}$ cm$^{-2}$ and 3x$10^{10}$ cm$^{-2}$. **Inset in (a):** Time evolution of the nonlinear trion signal, obtained by pumping at the exciton resonance, at 5 K and exciton density of 3x$10^{10}$ cm$^{-2}$. **Inset in (b):** Schematics of the trion transition. $E_{Trion}$(k) and $E_{electron}$(k) represent the trion and electron band dispersion, respectively. $f_{electron}$ represent the electron distribution.

In Fig. 2, we plot a set of differential reflectivity spectra, obtained when pumping selectively at the exciton resonance, at zero delay time between pump and probe pulses for different temperatures (Fig. 2a) and for different exciton densities (Fig. 2b). These differential spectra all evidence a clear blue-shift of the exciton line. This renormalization of the exciton resonance has been extensively studied in undoped-semiconductor quantum wells [13-16] and, is attributed to short-range exchange [14] having its origin in the Pauli exclusion-principle acting on the Fermi particles forming the excitons. This is a repulsive electron-electron and hole-hole interaction. It does not show up in 3D systems because it is compensated almost exactly by the attractive long-range correlations. The latter corresponds to the van der Waals attraction in the exciton gas and would induce a red-shift on the exciton resonance. The exciton blue-shift appears in two dimensional semiconductors [15] because the long-range Coulomb correlation effect is strongly reduced [14]. This means that, in our sample, the existence of the exciton blue-shift, due to the generation of an exciton population, is the signature that the long-range Coulomb correlation has an effect that we can neglect to first order compared to the more efficient short-range exchange interaction.

The novelty in our investigations is that the photo-generated exciton population induces also nonlinear effects on the trion resonance. We observe both a bleaching and a red-shift of the trion resonance, which depend on temperature (Fig. 2a) and exciton density (Fig. 2b). As evidenced in Fig. 2, both crossed effects already occur at zero delay time. The trion bleaching increases with excitation density and broadens with temperature. In insert of Fig. 2a, we show the time evolution of the nonlinear trion signal. It is important to note that the trion bleaching signal does not increase with time, as would be expected if this bleaching was only due to a trion population. However, at 5 K, the trion formation from an exciton population takes place within about 10 picoseconds [17]. Consequently, at short delay after excitation, only a small amount of trions (less than 10 %) is formed in the quantum well. The almost constant value of the trion bleaching over about 10 ps evidences that this bleaching evolves according to two opposite contributions of the same order of magnitude: the contribution due to the trion population which increases with time, and that due to the exciton population which decreases with time. Thus, we attribute the trion bleaching, at short times, to the phase-space filling of the optically-accessible k-space by the photo-generated excitons, resulting into a blocking of the trion transition. This result evidences that excitons and trions share the same k-space and originate from the same ground state.

In the inset of Fig. 2a, the red-shift of the trion line, i.e. an increase of the low energy edge absorption of the trion transition, can only be observed in the first 2.5 ps. This red-shift effect can not be attributed to a renormalization of the trion transition due to trion population because, as we will see bellow (Fig. 3), a trion population does not induce a red-shift in the trion resonance and at short times, there are only electrons and the photo-generated excitons in the quantum well. We suggest that the most likely origin of the red-shift effect is a change in the distribution of electrons with the presence of photo-generated excitons.

The resonant excitation that we are using does not allow sizeable heating of the electron gas. The possible absorption, by electrons, of acoustic-phonons emitted by excitons in the trion formation process is in a time scale of about 10 ps (trion formation time [17]). In figure 2b, we show that the red-shift of the trion resonance depends on the exciton density. We propose that the possible explanation for the change in the electron distribution is through the exciton-electron interaction. This interpretation is supported by the reported work of Honold et al [18], which demonstrated that the exciton dephasing time decreases with the presence of an incoherent electron population in the quantum well. In our experiment, we estimate the exciton dephasing time to be 1 ps or even less [19, 20].

Thus, as excitons are excited they interact with electrons inducing a broadening in the homogeneous line of excitons as well as of that of electrons. This results in an increase of the electron occupation at higher energies favouring the trion transitions at lower energies (diagram in Fig. 2b). The trion red-shift is then observed at short times and, at later times, it is masked by the bleaching effect due to trion formation from the exciton population. Previous investigations [14, 21] give strong evidence that the changes in the exciton absorption in the presence of carriers in the quantum well are dominated by phase-space filling and exchange. The long-range Coulomb interaction was shown to be a less important effect. These previous studies bring us to propose that excitons interact with electrons by short-range fermion exchange.

At higher temperatures, 10 K and 20 K, the red-shift of the trion resonance is not observed. At these temperatures, the homogeneous broadening of the electron distribution is essentially due to electron–acoustic phonon scattering, which is the dominant electron scattering process in this temperature range [9]. Therefore, the exciton-electron interaction does not modify enough the electron distribution to induce an increase in the trion absorption at low energies. As a result, at higher temperatures, when excitons are created we only observe a broad bleaching of the trion resonance (Fig. 2a). This bleaching signal is the consequence of phase-space filling mainly by the excitons, at short times and, at longer times, is also attributed to trions while the thermal equilibrium between trion and exciton populations is reached [17].

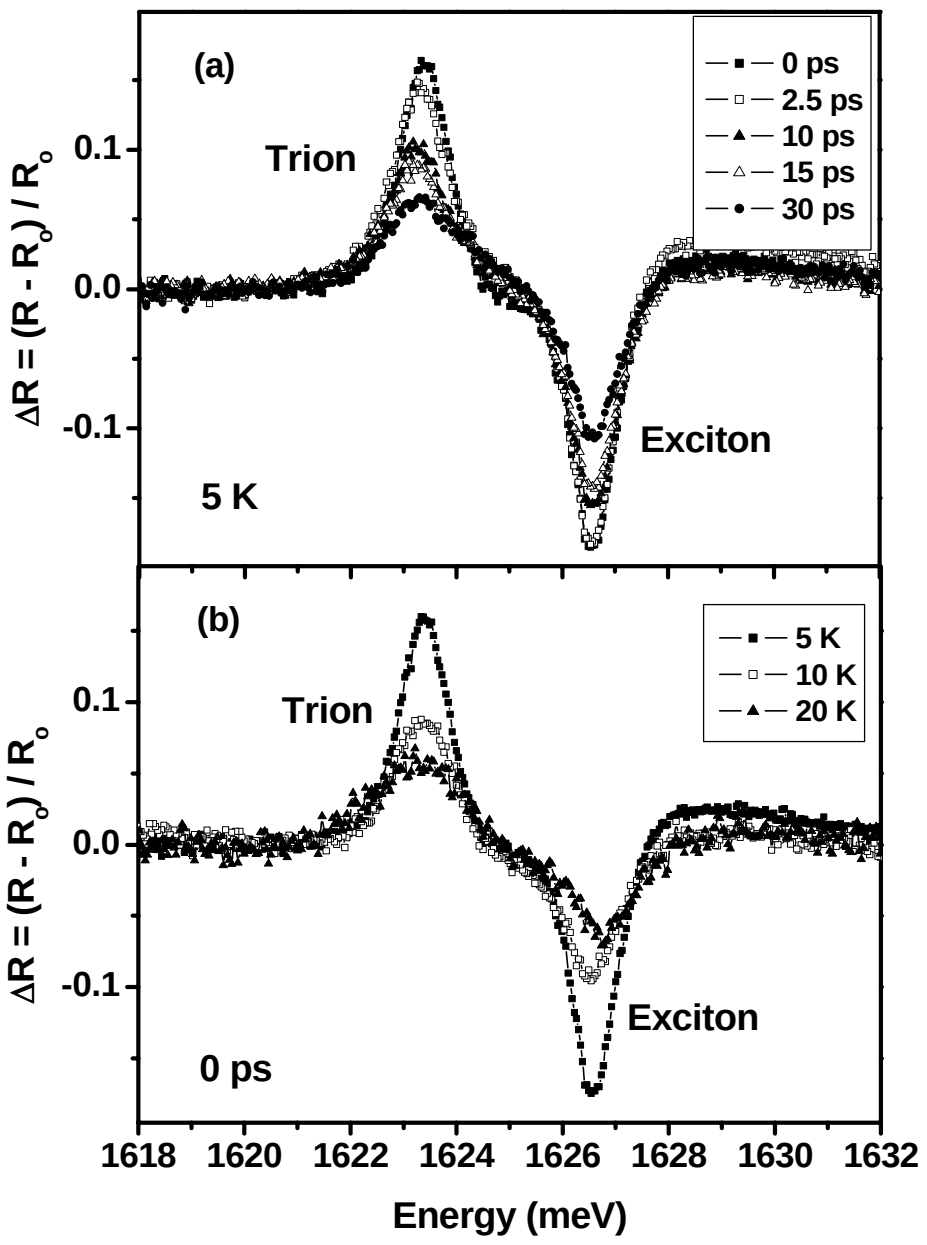

**FIG. 3:** Differential reflectivity spectra obtained by pumping at the trion resonance and with a trion density of about $2x10^{10}$ $cm^{-2}$: **(a)** at 5 K, $E_P$ = 1623.35 meV, and at different delay times and **(b)** at temperatures of 5 K ($E_P$ = 1623.35 meV), 10 K ($E_P$ = 1623.4 meV) and 20 K ($E_P$ = 1623.4 meV), at zero delay time.

After showing that the creation of an exciton population induces large effects on the trion resonance, let us turn to the effect of a trion population on the trion and exciton resonances. The nonlinear optical effects in the reflectivity spectrum for resonantly created trions are notably different from those induced by an exciton population (Fig. 3). The trion resonance is bleached without any shift in energy and we find an induced absorption of the exciton line. Figure 3a displays the time evolution of the differential reflectivity spectrum when the sample is excited at the trion resonance, at 5 K. In order to evaluate these trion and exciton oscillator strength variations, the spectrally integrated bleaching signal of trions and the spectrally integrated induced absorption signal of excitons are divided by 1.2 and by 1.6 (factors calculated with matrix transfer, see above), respectively. In Figure 4, they are plotted as a function of time, for several excitation powers. These results show that: (a) as trions are created in the well, the induced absorption gained by excitons reaches similar amplitude as the bleaching experienced by trions and (b) the induced absorption of excitons decays with trion bleaching. These evidence correlated effects acting on both signals. The bleaching of the trion line obviously originates from phase-space filling of trions. The induced absorption on excitons is not an expected effect.

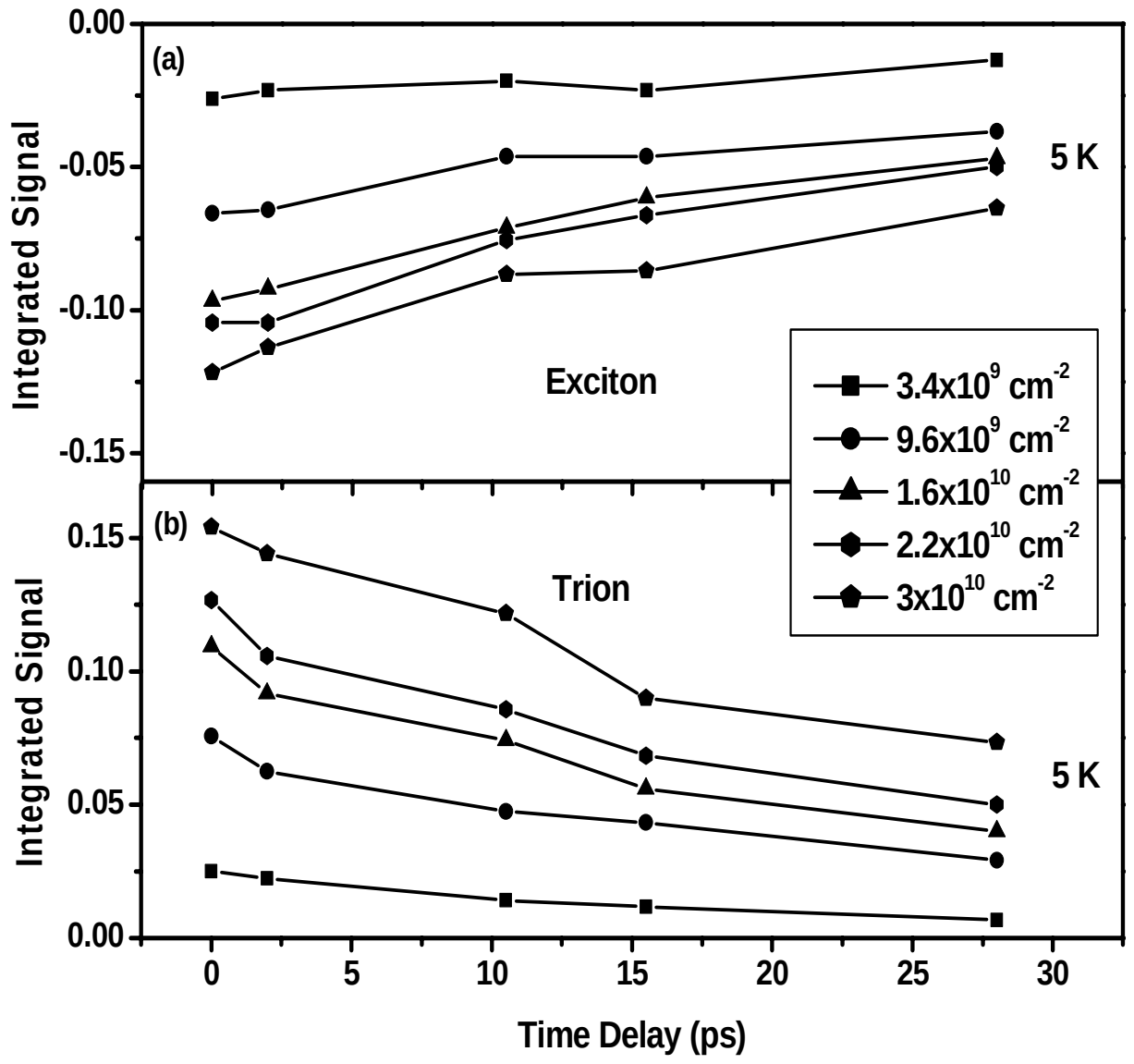

**FIG. 4:** **(a)** Integrated induced absorption signal of excitons and **(b)** integrated bleaching signal of trions, as a function of delay time for several excitation powers at the trion resonance, at 5 K.

It is known that, in undoped quantum wells, the oscillator strength of excitons is very large and, in modulation-doped quantum wells, the exciton loses its oscillator strength, due to the presence of an electron gas [3]. The reduction of the exciton oscillator strength in the presence of carriers has been attributed to the Pauli exclusion principle, due to both phase-space filling and "screening" due to short-range exchange interaction of the electron gas and, the long-range Coulomb screening has been considered to be a much weaker effect [14, 21, 22]. The exciton blue-shift, obtained from excitation at the exciton resonance, indeed evidences that the long-range Coulomb correlation may be neglected in our sample, which validates that the reduction of excitons oscillator strength cannot result from such correlations. However, in the same way excitons have been observed to block the trion transition due to phase-space filling, the presence of trions should block the exciton transition by the same phase-space filling argument. Thus, it is expected that the presence of trions, as well as the presence of electrons, should induce a loss in the exciton oscillator strength. In our experiment we obtain on the contrary an increase of the excitonic absorption when the electron population is transformed into a trion population. It is then necessary to invoke another effect to explain the increased absorption of excitons in our experiments. We attribute the gain of exciton oscillator strength, when trions are created, as resulting from the changes in the screening of excitons via short-range electron-electron exchange. It is worth noting that the exchange electron-electron interaction produces effects similar to that of the classical Coulomb interaction: the former also produces an “exchange hole”, by repelling all the electrons in the same spin state, just like the “correlation hole” in the latter [1]. Therefore, this extra “positive charge” generated from electron exchange process screens the exciton Coulomb potential. As trions are photo-generated, there are less electron in the electron gas to screen the exciton and the exciton gains back some oscillator strength. As a trion recombines, the electron is released back to the electron gas and is able again to screen excitons. Thus, we observe a recovery of the induced absorption signal of excitons with time (Figs. 3a and 4). This result reveals that electrons interact much more efficiently among themselves than trions do, which suggests that excitons are more dynamically screened by electrons than by trions. We would like to stress that the similar amplitude of the signal changes of the trion and exciton resonances, comes from two distinct effects, which are both the consequence of Pauli exclusion principle: phase-space filling for trions and screening reduction for excitons. This is an intriguing result and we believe that only a theoretical description can explain it, which is beyond the scope of this paper.

The above explanation in terms of more efficient screening of excitons by electrons than by trions is further corroborated by the absence of an energy shift of the trion resonance upon resonant pumping, which evidences the ineffectiveness of the trion-trion interaction. Note that, exciton-exciton scattering is more effective since it induces a blue-shift in the exciton resonance (Fig. 2). Therefore, we are led to assert that exciton-exciton exchange interaction is more efficient than that of trion-trion interaction. The elastic exciton-exciton Coulomb scattering in semiconductor quantum wells, following a resonant exciton excitation, has been considered theoretically by Ciuti et al [23]. They have shown that electron-electron and hole-hole exchange terms are the dominant ones, being much larger than the classical electrostatic dipole-dipole interaction as well as the direct exciton-exciton exchange. Furthermore, their computation shows that electron-electron and hole-hole exchange terms give equal contributions to the interaction process. In a simple minded approximation, we could expect that the inter-trion exchange interaction between carriers could be larger than that of inter-exciton exchange interaction, because trions are formed by three particles instead of two as excitons. In our knowledge, there is no theoretical investigation on the short-range exchange between trions, but our results suggest that the trion-trion exchange interactions are less efficient than the exchange interaction between excitons. The localization of trions may play an important role in this inefficient inter-trion exchange process. Trions have been indeed found to be more localized than excitons, at low temperature [19].

At higher temperatures, trions are no more localized [19] and thus, we could expect, when a trion population is generated, a blue shift of the trion line. We have performed experiments at 10 K and 20 K to test this premise (Fig. 3b). We found that there is no blue-shift of the trion resonance. Additionally, the induced absorption of excitons persists as a correlated effect between excitons and trions [24]. This result shows that trion-trion interaction is always inefficient whether they are localized or not. As the temperature is increased, the main scattering process of trions is with acoustic-phonons via deformation potential, which is much more efficient than the exciton- (and electron-) acoustic phonon scattering because of the larger interaction potential of trions [19]. This is evidenced by the broadening of the trion resonance with temperature (Fig. 3b). Both results together demonstrate, without ambiguity, that the trion-trion interaction is not the dominant scattering process for trions.

In conclusion, novel results were presented that reveal how the nonlinearities induced by trions are different from those induced by excitons, and in addition they are mutually correlated. We attribute the main source of these distinct differences to Pauli exclusion-principle, which is the origin of phase-space filling and short-range fermion exchanges. We have observed phase-space-filling effect for resonantly created excitons as a bleaching of the trion transition, while an induced absorption of excitons is detected as trions are excited on resonance. The latter result strongly suggests that excitons are more dynamically screened by free electrons than by trions. For resonantly created excitons, a blue shift of the excitonic transition is observed due to repulsive exciton-exciton interactions while no energy shift was seen at the trion line when trions are generated on resonance. This result evidences the significance of the exciton-exciton and trion-trion interactions. On the other hand, as excitons are created, the trion resonance undergoes a red-shift, which we attribute to short-range exciton-electron interactions. Our experimental results clearly reveal crossed and correlated effects on trion and exciton resonances. This yields new insight on many-body effects that should stimulate theoretical studies of interactions of excitonic complexes taken into account the presence of electrons, excitons and trions.

This work was partially supported by Swiss National Science Foundation. We are grateful to Cristiano Ciuti and Daniel Oberli for enlightening discussions and for critical reading of the manuscript. We wish to thank M. Combescot and R. Zimmermann for helpful discussions.

---

* Present address: IQUEST, University of California, Santa Barbara, California 93106, USA.

be properly lorentzian because of the recoil of the electron [A. Esser, R. Zimmermann, E. Runge, Phys. Stat. Sol. (b) **227**, 317 (2001)] that broadens the resonance. However, this effect is masked by the inhomogenous broadening (about 1meV) [9] that we have added to account for the quantum well inhomogeneities [L. C. Andreani et al., Phys. Rev. B 57, 4670 (1998)].

[20] For low exciton density of $3x10^{9}$ $cm^{-2}$ and electron concentration ($2x10^{10}$ $cm^{-2}$) the dephasing time is 10 ps at 2 K [ref. 19]. For doubled density of excitons ($6x10^{9}$ $cm^{-2}$) the dephasing time is 8 ps. It has been reported that the interaction between excitons and electrons can be 8 to 10 times stronger than between excitons [ref. 18]. Therefore, the exciton dephasing time for an electron concentration of $4x10^{10}$ $cm^{-2}$ can be estimated to be 1 ps or even less.